\documentclass[10pt]{iopart}

\usepackage{iopams}  
\usepackage{graphicx}
\usepackage{url}
\usepackage{booktabs}

\graphicspath{{figures/}}

\def\ket#1{|\,#1 \,\rangle}

\def\ms{\,\mathrm{ms}}

\def\MHz{\,\mathrm{MHz}}

\def\eref#1{(\ref{#1})}
\def\fref#1{figure~\ref{#1}}
\def\tref#1{table~\ref{#1}}

\begin{document}

\title[Absolute frequency measurement of a Lu\textsuperscript{+} (\textsuperscript{3}D\textsubscript{1}) optical frequency standard ]{Absolute frequency measurement of a Lu\textsuperscript{+} (\textsuperscript{3}D\textsubscript{1}) optical frequency standard via link to international atomic time}


\author{$^1$Zhao Zhang, $^1$Qi Zhao, $^1$Qin Qichen, $^1$N. Jayjong, $^1$M. D. K. Lee, $^1$K. J. Arnold, $^{1,2}$M. D. Barrett}
\address{$^1$Centre for Quantum Technologies, National University of Singapore, 3 Science Drive 2, 117543 Singapore}
\address{$^2$Department of Physics, National University of Singapore, 2 Science Drive 3, 117551 Singapore}
\ead{cqtkja@nus.edu.sg}
\vspace{10pt}

\begin{abstract}
We report on an absolute frequency measurement of the ${\rm Lu}^{+}\,(^{3}\rm D_1)$ standard frequency which is defined as the hyperfine-average of $^{1}\rm S_0$ to $^{3}\rm D_1$ optical clock transitions in $^{176}{\rm Lu}^{+}$. The measurement result of $353\,638\,794\,073\,800.35(33)$Hz with a fractional uncertainty of  $9.2 \times 10^{-16}$ was obtained by operating a single-ion $^{176}{\rm Lu}^{+}$ frequency standard intermittently over 3 months with a total uptime of 162 hours. Traceability to the International System of Units (SI) is realized by remote link to International Atomic Time. This is the first reported absolute frequency value for a ${\rm Lu}^{+}\,(^{3}\rm D_1)$ optical frequency standard.
\end{abstract}

%
\vspace{2pc}
\submitto{\MET}

%
\maketitle
%
\ioptwocol
\section{Introduction}
The International System of Units (SI) defines the second by the duration of a fixed number of oscillation periods of the caesium ($^{133}\rm Cs$) ground-state hyperfine transition. The best primary frequency standards (PFS) to realize the SI second are Cs fountains with an inaccuracy near one part in $10^{-16}$ ~\cite{weyers2018advances,beattie2020first}.  Several frequency standards based on optical transitions of different ions and atoms now report uncertainties near one part in $10^{-18}$ ~\cite{mcgrew2018atomic,takamoto2020test,sanner2019optical,aeppli2024clock,brewer2019al+,zhiqiang2023176lu+,hausser2025in+} and a roadmap to the redefinition of the SI second has been devised~\cite{dimarcq2024roadmap}. The International Committee for Weights and Measures (CIPM) already maintains a list of recommended values for standard frequencies, many recognized as secondary representations of the second (SRS). Some optical clocks based on SRS reference transitions are officially recognized as secondary frequency standards (SFS) and in recent years contribute to the steering of International Atomic Time (TAI) through reporting of measurements to the International Bureau of Weights and Measures (BIPM).   For a recently established optical clock species such as Lu$^+$, absolute frequency measurements are necessary for CIPM to establish a recommended value, which are prerequisite to becoming a SRS.

\begin{figure}[t]
   \centering
   \includegraphics[width=\columnwidth]{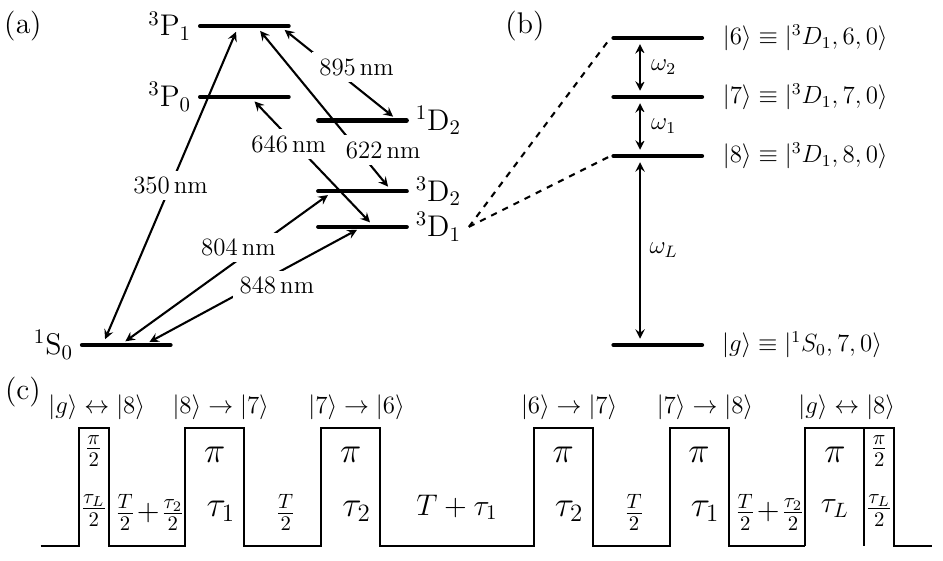}
   \caption{(a) Energy level structure of $^{176}\mathrm{Lu}^{+}$. (b) Microwave and optical transitions used in the clock interrogation sequence. (c) Clock interrogation sequence for hyperfine averaging with hyper-Ramsey spectroscopy.}
   \label{fig:Lulevel}
\end{figure}

The $^{1}\rm S_0$ to  $^{3}\rm D_1$ transition of $^{176}{\rm Lu}^{+}$ is a promising reference for which $\lesssim 10^{-18}$ uncertainty can be realized with relative ease~\cite{zhiqiang2023176lu+}. We report on the first absolute frequency measurement of the Lu\textsuperscript{+}(\textsuperscript{3}D\textsubscript{1}) optical standard. In section~\ref{sec:ExpSys}, we outline the $^{176}{\rm Lu}^{+}$ clock operation and the frequency measurement scheme to link to the SI second. In section~\ref{sec:sys} we detail the frequency shifts and uncertainties for the ${\rm Lu}^{+}$ clock. In section~\ref{sec:link}, we analyze the uncertainty contributions from each step of the frequency chain. Finally we discuss results in section~\ref{sec:conclusion}. 

\begin{figure*}[ht]
    \centering
    \includegraphics[width=0.7\textwidth]{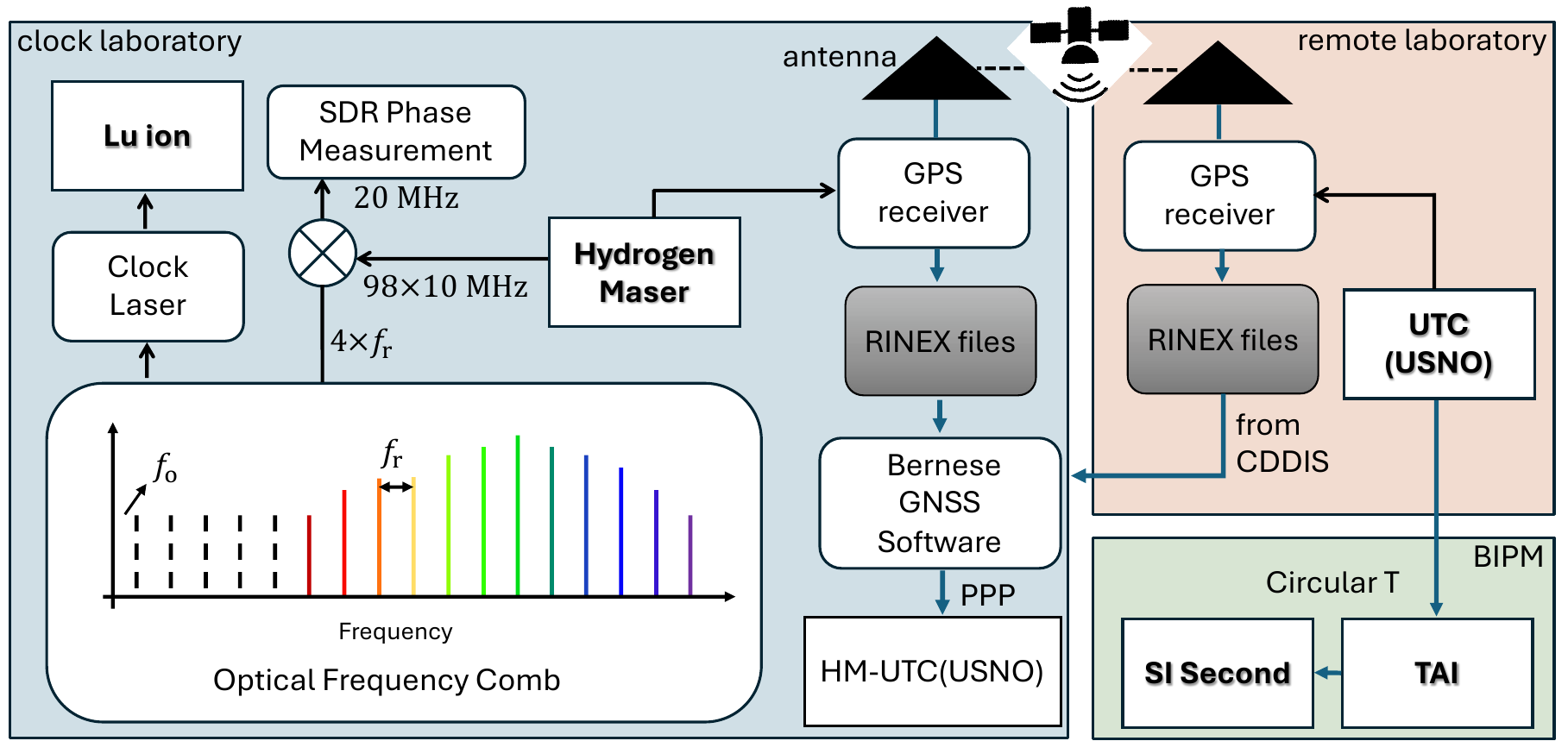}
    \caption{Remote comparison of the lutetium clock to international atomic timescales for absolute frequency evaluation. The optical clock is compared to a hydrogen maser which serves as the local flywheel. The hydrogen maser is linked to the timescale of a remote laboratory, UTC(USNO), via GPS link. UTC(USNO) is linked to TAI and the SI second through the circular T reports of BIPM.}
    \label{fig:si_link}
 \end{figure*}
 
\section{Experimental System and Methods \label{sec:ExpSys}}
\subsection{$^{176}{\rm Lu}^{+}$ optical frequency standard}
Experiments are performed with a single $^{176}\mathrm{Lu}^{+}$ ion in a linear Paul trap denoted as Lu-2~\cite{zhiqiang2023176lu+,EMM2024}. Figure \ref{fig:Lulevel}(a) shows the energy levels and transitions relevant for laser cooling, trapping, and state manipulation. 

For $^{176}\mathrm{Lu}^{+}$ hyperfine averaging is used to realize an effective $J=0$ level which practically eliminates second-order Zeeman shifts, and shifts arising from rank 2 tensor interactions, such as the electric quadrupole moment~\cite{MDB1}. The reference frequency, $f_\mathrm{Lu}$, of the ${\rm Lu}^{+}\,(^{3}\rm D_1)$ standard is thus defined as the average of the three optical transitions from $\ket{g}\equiv\ket{^1S_0,7,0}$ to the hyperfine states $\ket{F}\equiv\ket{^3D_1,F,0}$ where $F=6,7,$ and $8$. This average can be conveniently realized using a hybrid microwave and optical interrogation~\cite{kaewuam2020hyperfine}.  As illustrated in \fref{fig:Lulevel}(b-c), the interrogation sequence consists of an optical Ramsey interrogation on the $\ket{g}\leftrightarrow\ket{8}$ transition, where additional microwave pulses transfer population between the hyperfine states during the interrogation time.  Their timing is chosen such that the effective time in each of the hyperfine states is the same.  When the laser is servoed to the central fringe of the resulting Ramsey spectrum, the hyperfine-averaged (HA) clock frequency is given by 
\begin{equation}
f_\mathrm{Lu} = \frac{1}{3}(\nu_{8}+\nu_{7}+\nu_{6})=\nu_L+\frac{1}{3}(2f_1+f_2),
\label{Eq:clockfreq}
\end{equation}
where $\nu_{F}$ is the resonant frequency of the $\ket{g}\leftrightarrow\ket{F}$ optical transition, $\nu_L=\omega_L/(2\pi)$ is the frequency of the probe light, and $f_k=\omega_k/(2\pi)$ for $k=1,2$ are the microwave frequencies used within the interrogation sequence.  Even if the microwave fields are not exactly resonant, the interrogation sequence ensures $\nu_L+(2f_1+f_2)/3$ represents the HA clock frequency, as indicated by (\ref{Eq:clockfreq})~\cite{zhiqiang2023176lu+,kaewuam2020hyperfine}.

Each experiment starts with Doppler cooling on the 646 nm transition followed by preparation in the $\ket{^3D_1,8,0}$ state and then the Ramsey interrogation sequence as shown in \fref{fig:Lulevel}(c). The timing parameters used for the Ramsey interrogation are $\tau_\mathrm{L} = 1.4\,\ms$, and $\tau_1=\tau_2=2\,\ms$ with total Ramsey interrogation time $T_\mathrm{R} = 3(T+\tau_1+\tau_2) = 42\,\ms$. The additional 848-nm $\pi$-pulse before the end of the sequence shown in \fref{fig:Lulevel}(c), is phase shifted by 180$^\circ$ and included for suppressing the remaining probe light induced ac Stark shift by the hyper-Ramsey method~\cite{yudin2010hyper,hyperRamsey2012peik}. A frequency discriminator for locking to the central Ramsey fringe is generated by repeating the entire sequence twice per cycle, alternating the phase shift on the first $\pi/2$-pulse by $\pm{90}^\circ$ ~\cite{hyperRamsey2012peik}.  Every $N=25$ cycles, which takes $t_u = 3.6$ seconds, the 848 nm laser frequency is steered to maintain the condition given by \eref{Eq:clockfreq}. For the Ramsey time used here, clock interrogation accounted for 60\% of the duty cycle, with the remaining time allocated to cooling (10 ms), state preparation (5 ms), and detection (1 ms).

The microwave frequencies $f_1$ and $f_2$ remain fixed throughout all measurements. All synthesized rf offsets and the microwave frequencies are referenced to the hydrogen maser (Microchip MHM-2010). 

The clock laser near 848 nm is an external cavity diode laser which is stabilized to a 10-cm ultra low expansion (ULE) optical cavity with finesse of 400000. This cavity is under vacuum at $1\times10^{-7}$ mbar and actively temperature stabilized to 33.2 $^\circ$C, corresponding to its zero coefficient of thermal expansion point.  A wideband fiberized electro-optic modulator (EOM) is used to frequency offset the clock laser from the optical reference cavity.  The radio-frequency (rf)  driving the EOM is generated by a direct digital synthesizer (DDS) which both continuously compensates the long term linear creep of the ULE cavity of 40 mHz s$^{-1}$ and applies the phase-continuous frequency corrections when servoing to the atomic reference.  An instability of $1.5\times10^{-15}$ at 1$\,$s was measured in a three-corner hat measurement against two additional optical reference cavities.  

The choice of Ramsey time and number of cycles between servo updates is limited by the clock laser stability. For this work the Ramsey time was chosen conservatively to ensure robust operation in which the servo never slips to an incorrect Ramsey fringe. The quantum projection noise limited fractional frequency instability during clock operation is
\begin{equation}
\sigma(\tau) = \frac{1}{f_\mathrm{Lu}}\frac{1} { 2\pi C T_R \sqrt{2 N}}\sqrt{\frac{t_u}{\tau}} \approx \frac{3.9 \times 10^{-15}}{\sqrt{ \tau}},
\end{equation}
for averaging time $\tau\gg t_u$ (in seconds). Here $C\approx0.75$ is the Ramsey fringe contrast at $T_R=42$ ms.  

As illustrated in \fref{fig:si_link}, the optical clock laser frequency is compared to the hydrogen maser (HM) reference frequency, $f_{\mathrm{HM}} =10\,\MHz$, using a commercial Er-doped femtosecond fiber frequency comb (Menlo Systems). The $\sim$250 MHz repetition rate ($f_\mathrm{r}$) is referenced to the $^{176}\mathrm{Lu}^{+}$ stabilized clock laser by phase locking the beat note between the laser and comb to a DDS. The carrier-envelope offset frequency of the comb, $f_\mathrm{o}$, is phase locked to the HM. The HM signal is multiplied up to $980\,\MHz$ using a phase locked dielectric resonator oscillator and mixed with the fourth harmonic of $f_r$ at 1 GHz. The phase of the down mixed signal at 20 MHz is measured continuously with zero dead-time using software defined radio (SDR) hardware (USRP N210)~\cite{sherman2016oscillator}. No cycle slips of the frequency comb were observed during the optical clock uptime. The SDR measurement system contributes an uncertainty of $3\times10^{-14}\, (\tau/\mathrm{s})^{-1}$ to the measurement of $\frac{f_\mathrm{Lu}}{f_\mathrm{HM}}$ for averaging time $\tau$.  

\subsection{Link to SI second}
The system to link the optical clock to the SI second is outlined in \fref{fig:si_link}.  Traceability to the SI second is achieved by a link to International Atomic Time (TAI), which differs from Coordinated Universal Time (UTC) by an integer number of leap seconds, but shares the same scale interval. UTC is a virtual timescale with local realisations, UTC($k$), at laboratories reporting to BIPM.  Since a local realization is not directly available in our lab, we rely on a GPS link to a UTC($k$) realized in a remote laboratory~\cite {Petit2018}. Specifically we choose UTC(USNO) because it has good stability and GPS receiver data is freely available.  The data from GPS receiver USN7, which is referenced to UTC(USNO)~\cite{jiang2017long} is obtained from CDDIS~\cite{USNOdata,NOLL20101421} in Receiver Independent Exchange (RINEX) format. Local RINEX files are generated by a dual-band GNSS receiver (PolaRx5TR) referenced to the HM.  A Precise Point Positioning (PPP) algorithm~\cite{Petit2009,Ray_2005} is implemented using the Bernese GNSS software~\cite{Bernese} to evaluate the phase difference $\Delta x = x(\mathrm{HM})-x(\mathrm{USNO})$ on 30 s intervals.   As shown in \fref{fig:maser_allan} (orange), the PPP link instability is observed to average down as $4.6\times10^{-15}\,(\tau/\mathrm{day})^{-0.75}$ until reaching a $1.3\times10^{-15}$ frequency flicker noise floor attributed to the HM at around 5 days.  The PPP result is used to evaluate the HM frequency offset and drift referenced to UTC(USNO) over an 80 day interval encompassing the optical clock measurements. The HM acts as a flywheel for interpolation between optical clock operation.

Circular T~\cite{circulart} data published by BIPM links UTC(USNO) to UTC, and thereby TAI, on five day intervals. TAI is a realisation of terrestrial time (TT), which is a coordinate time scale defined in the geocentric reference frame with the scale unit SI second on the geoid. The estimated deviation between TAI and TT is given by $y_\mathrm{TAI} = -d$ where $d$ is published in part 3 of the Circular T report for each month. 

The complete frequency chain is summarized as
\begin{eqnarray}
    \frac{f_\mathrm{Lu}}{f_\mathrm{SI}}&=\frac{f_{\mathrm{Lu}, T_1}}{f_{\mathrm{HM}, T_1}}\times\frac{f_{\mathrm{HM}, T_1}}{f_{\mathrm{HM}, T_2}}\times\frac{f_{\mathrm{HM}, T_2}}{f_{\mathrm{UTC(USNO)}, T_2}}\nonumber\\ 
    &\times\frac{f_{\mathrm{UTC(USNO)}, T_2}}{f_{\mathrm{TAI}, T_2}}\times\frac{f_{\mathrm{TAI}, T_2}}{f_{\mathrm{TAI}, T_3}}\times\frac{f_{\mathrm{TAI}, T_3}}{f_{\mathrm{SI}, T_3}} \label{eq:link}
\end{eqnarray}
where $T_1$ corresponds to the uptime intervals of the Lu$^+$ frequency standard, $T_2$ and $T_3$ are the 5-day and 1-month intervals aligned to the Circular T respectively.  The ratios ${f_{\mathrm{HM}, T_1}}/f_{\mathrm{HM}, T_2}$ and ${f_{\mathrm{TAI}, T_2}}/{f_{\mathrm{TAI}, T_3}}$ account for the uncertainty in interpolation of the HM and TAI respectively.


\begin{table}[t]
\centering
\footnotesize

\caption{\label{lusys} Nominal uncertainty budget of the Lu$^+$ clock for operating parameters used in this campaign. All shifts and uncertainties are given fractionally relative to the 353 THz ${\rm Lu}^{+}\,(^{3}\rm D_1)$ standard frequency. }

\lineup
\begin{tabular}{lrr}
\toprule
& Shift  & Uncertainty  \\
Effect &($\times10^{-18}\,\mathrm{Hz/Hz}$) &($\times10^{-18}\,\mathrm{Hz/Hz}$)\\ \midrule
Quadratic Zeeman shift&-140.7&0.6\\
Excess micromotion&-0.2&0.2\\
Second-order Doppler &-0.1&0.1\\
ac-Zeeman (rf)&0.2&$<$0.1\\
ac-Zeeman (microwave)&-8.0&6.2\\
Microwave coupling&0.0&3.7\\
ac-Stark (848 nm)&0.0&1.2\\
Blackbody radiation&-1.6&0.3\\
\midrule
Total &-150.4&7.3\\
Gravitational redshift &1718\,\,\,\,&36\,\,\,\,\\
\midrule
Total (with redshift) &1568\,\,\,\,&37\,\,\,\,\\
\bottomrule
\end{tabular}
\end{table}

\section{Lu$^+$ Systematics \label{sec:sys}}
The Lu$^+$ clock systematic uncertainty budget for the operating parameters of this measurement campaign is summarised in \tref{lusys}. Data presented in this work was collected using the apparatus `Lu-2' shortly after the clock comparison and systematics evaluation reported in \cite{zhiqiang2023176lu+}. Compared to the uncertainty budget in  \cite{zhiqiang2023176lu+}, probe induced systematic shifts are larger here due to the shorter Ramsey interrogation time.  In \cite{zhiqiang2023176lu+} two atomic references were compared using correlation spectroscopy which enabled interrogation times beyond the coherence time of the laser.  The only frequency corrections $>10^{-17}$ and of statistical significance for the absolute frequency assessment are from the quadratic Zeeman shift and gravitational redshift. Remaining systematics are only briefly discussed.

\subsection{Quadratic Zeeman}
A residual quadratic Zeeman shift remains after HA due to hyperfine-mediated mixing between fine structure levels~\cite{zhiqiang2020hyperfine}. The quadratic Zeeman coefficient has been measured to be -4.89264(88) Hz/mT$^2$ \cite{zhiqiang2023176lu+}. Measurement of the linear Zeeman splitting on the $\ket{^3D_1,6,0} $ to $\ket{^3D_1,6,\pm1}$ microwave transition is regularly interleaved during clock operation from which the magnetic field at the ion is inferred via the g-factor $g_6 = -0.071\,662\,506(33)$ reported in~\cite{zhiqiang2020hyperfine}. There is no magnetic shielding or any steering of the applied magnetic field. The mean and standard deviation of all magnetic field measurements over the full campaign was 0.10084(20) mT.  The corresponding quadratic Zeeman shift is -49.75(20) mHz, or $-1.407(7) \times10^{-16}$ fractionally, where the uncertainty is determined by the standard deviation of the field measurements. We note that for highest accuracy operation the magnetic field can be monitored continuously to correct the quadratic Zeeman shift on a point by point basis. In this case the uncertainty of the quadratic Zeeman evaluation is limited only by the measured quadratic Zeeman coefficient to $2.5\times10^{-20}$ at a typical operating field of 0.1 mT assessed to 100 nT uncertainty. 

\begin{figure}[t]
    \centering
    \includegraphics[width=\columnwidth]{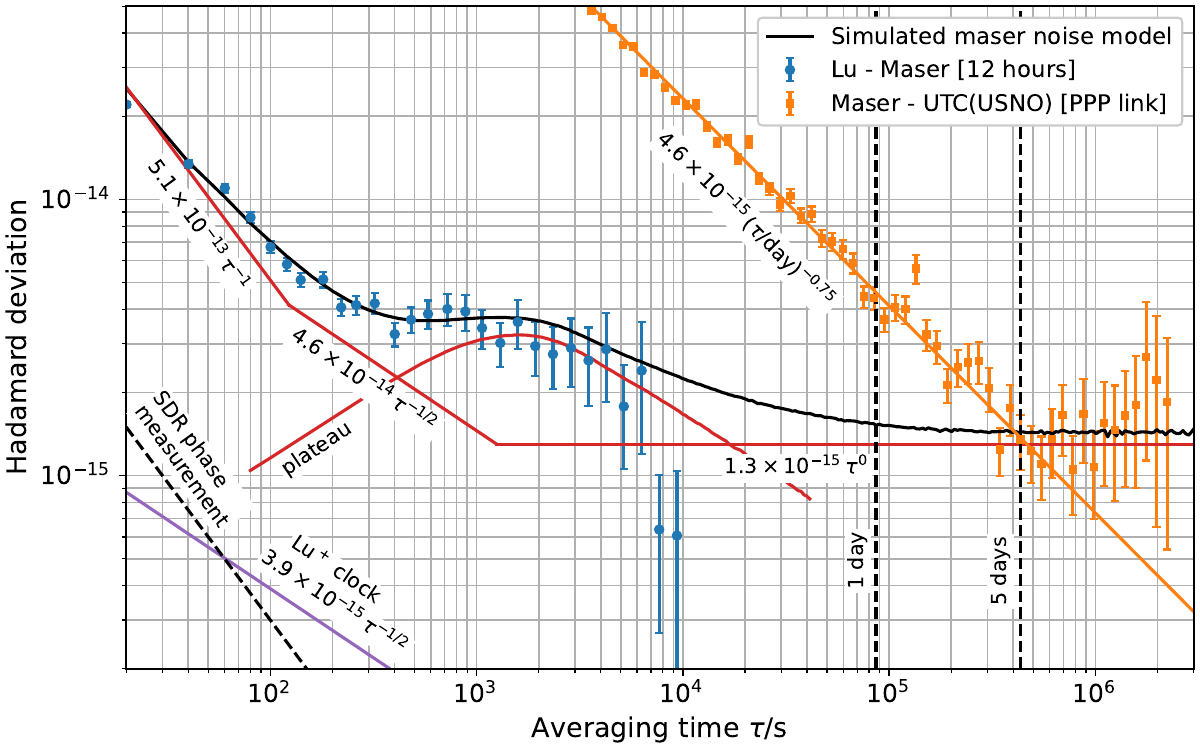}
    \caption{HM instability observed relative to the Lu$^+$ clock from 12 hours continuous measurement (blue circles) and relative to UTC(USNO) via the PPP link (orange squares). The black line is the numerical implementation of the noise model for the HM that is generated as the sum of noise processes for which the individual contributions are indicated by the red lines.}
    \label{fig:maser_allan}
 \end{figure}
 
\subsection{Gravitational redshift}

The redshift is given by $ \frac{\Delta f}{f}=\frac{g h}{c^2}$ where $g =  9.7803(3)\,\mathrm{m}~\mathrm{s}^{-2}$ is the local gravitational acceleration and the geoid height $h = H - N$ is obtained from the difference in the orthonormal height $H$ and geoid undulation $N$.  The orthonormal height of the trapped Lu$^+$ ion from the earth ellipsoid (WGS84) was measured to be $H=23.68(10)$ m by orthometric height leveling relative to the rooftop GNSS antenna~\cite{denker2018geodetic}. A geoid undulation of $N=7.89(10)$ m is obtained from the Earth Gravitational Model (EGM2008). The gravitational redshift is estimated to be 1.72(4)$\times 10^{-15}$, including additional uncertainty to allow for bias due to tidal modulation~\cite{muller2018high}.

\subsection{Other Systematics}
Excess micromotion was measured in three orthogonal directions on three occasions throughout the measurement campaign using sideband spectroscopy~\cite{berkelandMicro} on the 804 nm $\ket{^1S_0,7,0} $ to $\ket{^3D_2,9,0}$ clock transition. In each instance the magnitude of the corresponding second-order Doppler shift was $<2\times10^{-19}$ before (re-)optimizing the compensation.  It is noted that an improved micromotion compensation method was developed after this work~\cite{EMM2024}.

The Second-order Doppler shift due to thermal motion and the ac-Zeeman shift due to rf fields at the trap drive frequency are as evaluated in~\cite{zhiqiang2023176lu+}. The microwave ac-Zeeman shift arises from imbalanced $\sigma^\pm$ polarisation components of the microwave fields which off-resonantly couple to $\Delta m = \pm 1$ Zeeman transitions.  The polarisations of both microwave fields were evaluated three times throughout the campaign by measuring the relative coupling strengths of $\Delta m = [-1,0,1]$ microwave transitions. The observed coupling strengths did not vary by more the 1\% indicating stable microwave polarizations throughout.  One of the microwave fields had imbalanced $\sigma^\pm$ components resulting in an uncharacteristically large shift and uncertainty, $-8.0(6.2)\times10^{-18}$, due to this effect. 

An error in microwave coupling strengths relative to the values determined by their respective $\pi$ time gives rise to imperfect hyperfine averaging~\cite{zhiqiang2020hyperfine}. The uncertainty in \tref{lusys} corresponds to a 1\% error in the microwave couplings. 

For hyper-Ramsey spectroscopy~\cite{hyperRamsey2012peik} the clock laser is frequency stepped by the evaluated ac-Stark shift during the clock interrogation pulses. The remaining clock shift after hyper-Ramsey, in Hz, is given by $\frac{2}{\pi T_R} \left(\frac{\Delta}{\Omega} \right)^3$ where $\Delta$ is the error in the frequency step and $\Omega = \pi/\tau_L$ is the optical coupling strength. The ac-Stark shift is evaluated to $\sim$1\% precision prior to clock operation by interleaved comparison of hyper-Ramsey and Rabi spectroscopy. Accounting for possible variation in laser intensity over the course of clock operation, we allow for 5\% error in the ac-Stark shift step in estimating the uncertainty for \tref{lusys}.

The  $^{1}\rm S_0$ to $^{3}\rm D_1$ transition has low sensitivity to blackbody radiation (BBR)~\cite{arnold2018blackbody,arnold2019dynamic} such that for a crude 35(10) $^\circ$C estimate of the temperature the BBR shift is only $-1.56(27)\times10^{-18} $.  

We note that the systematic uncertainty reported here is evaluated for the operating parameter of this campaign. These parameters were not optimised to reduce systematic uncertainty beyond the requirements of the absolute frequency measurement. The microwave ac-Zeeman shift can be suppressed by balancing the $\sigma^\pm$ polarisation components, or using a longer microwave $\pi$ time. The ac-Stark shift can be suppressed by using a longer optical pulse duration. With improved laser stability, the Ramsey time can be extended to reduce all probe induced systematics effects, which includes the microwave coupling, the microwave ac-Zeeman, and the (848 nm) ac-Stark effects.

\section{Analysis of link to SI second\label{sec:link}}
Measurements of the $^{176}$Lu$^+$ frequency standard relative to the local hydrogen maser, $f_\mathrm{Lu}/f_\mathrm{HM}$, were taken between the modified Julian dates (MJD) 59270 and 59338, spanning March to May in the year 2021. The total measurement uptime was 162 hours, where the average length of a single continuous run was 2.9 hours.   
The analysis follows equation \eref{eq:link} and is carried out in four steps. First the maser drift and frequency offset are evaluated by the GPS link and subtracted from the local measurements. Second the measurements are averaged over 5-day intervals aligned to the Circular T, accounting for the uncertainty due to interpolation over dead times using the hydrogen maser as a flywheel. Third, the 5 day averages are linked to TAI by the Circular T with weighted averages evaluated for each month. Five day intervals without optical frequency measurements are considered dead time and an interpolation uncertainty using TAI as the flywheel is estimated. Finally, monthly averages are linked to SI by the Circular T.

\begin{figure}[t]
    \centering
    \includegraphics[width=\columnwidth]{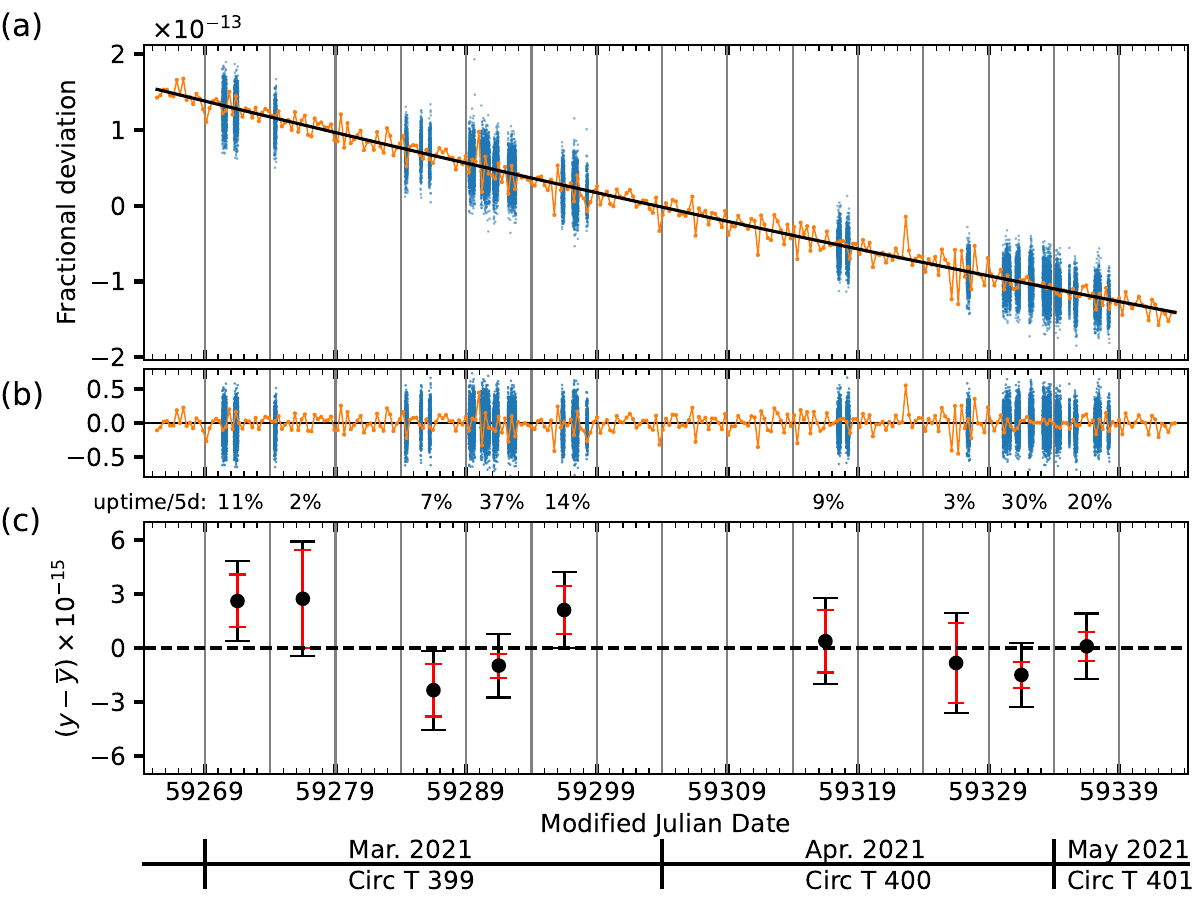}
    \caption{(a) Evaluation of the HM drift over an 80 day interval. The blue points are the intermittent measurements of the maser relative to the optical clock as logged at 20 second intervals. Orange points are the HM deviation measured against UTC(USNO) via the PPP link, plotted as 6 hour averages. A quadratic fit (black line) models the frequency offset and drift of the HM relative to UTC(USNO). (b) After subtraction of the fitted maser drift and offset, blue points are the optical measurements relative to UTC(USNO) and orange points are the HM-UTC(USNO) residual.  (c) Black points are the measurements $y[f_\mathrm{Lu}/f_\mathrm{TAI}]$ on 5-day intervals aligned to the Circular T reports. Outer black error bars represent the total statistical uncertainty, and inner red error bars are the uncertainty contribution only from dead time interpolation. The reduced chi-squared is $\chi^2_\nu = 0.67$ with $\nu=8$ degrees of freedom.   }
    \label{fig:measurements}
 \end{figure}
  
\subsection{Maser drift}
The hydrogen maser drift is evaluated relative to UTC(USNO) over an 80 day interval encompassing the optical frequency measurements, as shown in \fref{fig:measurements}. Over this interval the maser frequency drift slightly deviates from linear so a quadratic model is used.  The maser linear drift over this is interval is near $-3.7 \times 10^{-15}\,\mathrm{d}^{-1}$ and aging at a rate $4.6\times 10^{-15}\,\mathrm{d}^{-1} \mathrm{year}^{-1}$. The fitted maser drift model is subtracted from the local observations, the blue points in \fref{fig:measurements}(a), to link the measurements to UTC(USNO) up to uncertainty from the additional noise process on the maser frequency which are assumed to have zero mean. The blue points in \fref{fig:measurements}(b), which are $-y(\mathrm{Lu}/\mathrm{UTC(USNO)})$, are binned in 5 day intervals aligned to Circular T and averaged.  The uncertainty in the 5 d averages of $y(\mathrm{Lu}/\mathrm{UTC(USNO)})$ has three contributions.  First, the statistical uncertainty in the five day averages of $f_\mathrm{HM}/f_\mathrm{UTC(USNO)}$ relative to the maser drift model fit on the 80 day interval. This is estimated to be $1.5\times10^{-15}$ from the instability of the residuals, the orange points in \fref{fig:measurements}(b), at five days, and is consistent with the Hadamard deviation of $f_\mathrm{HM}/f_\mathrm{UTC(USNO)}$ which is shown in \fref{fig:maser_allan}. Second, the fitted drift model contributes uncertainty which is correlated across all measurements, and this is estimated to contribute $2\times10^{-16}$ to the final result. Third, the uncertainty contributed from interpolation of the HM over optical clock dead time, which is evaluated in the following section. 

\subsection{HM interpolation uncertainty}
The uncertainty due to HM interpolation over optical clock dead time is estimated by simulation of the maser noise~\cite{Pizzocaro_2020,Yu_2007}. A maser noise model is constructed to be consistent with both the observed $f_\mathrm{Lu}/f_\mathrm{HM}$ instability on timescales less than a few hours, and the maser flicker noise floor inferred from GPS link to UTC(USNO) on timescales longer than five days. The Hadamard deviation for the longest continuous measurement of $f_\mathrm{Lu}/f_\mathrm{HM}$ (12 h) is shown in \fref{fig:maser_allan}. The noise model is the sum of white phase noise $5.1\times10^{-13}\,(\tau/\mathrm{s})^{-1}$, white frequency noise $4.6\times10^{-14}\,(\tau/\mathrm{s})^{-1/2}$, flicker frequency noise $1.3\times10^{-15}$, and the additional noise labeled `plateau' in \fref{fig:maser_allan} which reproduces the instability plateau observed for averaging times in the interval $\sim$5-60 minutes. This plateau clearly cannot be modelled by only power law noise but is captured reasonably here with the addition of low-pass filtered white frequency noise.  The plateau has been consistently observed in $f_\mathrm{Lu}/f_\mathrm{HM}$ measurements over the last three years.  

\begin{figure}[t]
    \centering
    \includegraphics[width=\columnwidth]{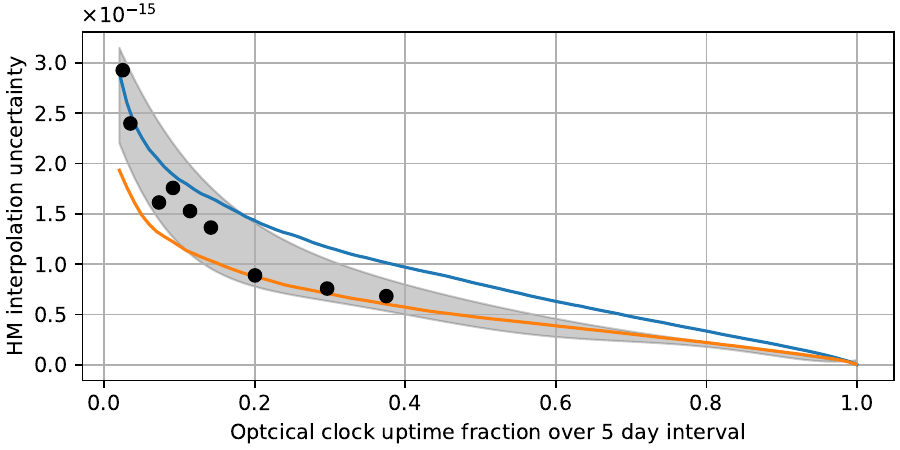}
    \caption{Result of simulation to estimate the uncertainty due to HM interpolation over optical clock dead time. Black points are uncertainties estimated by simulating the uptimes distributions present in the experimental data. See main text for description of other simulated results shown.}
    \label{fig:deadtime}
 \end{figure}

Power law noise is generated by fast Fourier transform (FFT) methods~\cite{timmer1995generating}. For the `plateau' noise, a high frequency cutoff is applied before the FFT to achieve low pass filtering. The total noise is generated by summing the time series generated for the individual processes.  The total noise is sampled to simulate $\sim2\times10^3$ trajectories on the interval $T_2= 5\,\mathrm{d}$. For an uptime $T_1$, which may be distributed non-continuously over the interval $T_2$, the dead time uncertainty is calculated from the standard deviation of the difference in averages over $T_1$ and $T_2$.  For the actual uptimes $T_1$ in the experimental data, the evaluated dead time uncertainties are shown as the black points in \fref{fig:deadtime}. The influence of the distribution of $T_1$ was further investigated by sampling random arrangements of $T_1$ as a function of total uptime fraction, with the restriction of a minimum (maximum) continuous experiment run time of 1 (12) hours to mimic typical operation. The gray shaded region in \fref{fig:deadtime} bounds the maximum and minimum dead time uncertainty found under these criteria. The solid lines correspond to two special cases, $T_1$ as a single continuous run centered on $T_2$ (blue) and $T_1$ divided into five runs evenly spaced over the five days (orange). For future measurement campaigns this simulation can serve as a practical guide to determine the uptime fraction and distribution of dead time required to achieve a target uncertainty.

\subsection{Diurnal effects}
Correlations with a daily temperature cycle are of particular concern because the majority of the optical clock uptime was during the daytime.   We analyzed all measurement residuals after maser drift subtraction for correlation with the time of day, but no definitive diurnal amplitude was observed. Statistically, we can only bound the amplitude to $< 1 \times 10^{-15}$ and at this upper bound the final result would be biased by $5\times10^{-16}$ when accounting for the distribution of uptime. We include this potential bias as a systematic uncertainty in the final result.

\subsection{Link to TAI}
The frequency ratio $f_\mathrm{UTC(USNO)}/f_\mathrm{TAI}$ is evaluated from the phase data reported on 5-day intervals in the Circular T reports. The recommended link uncertainty for satellite frequency transfer is~\cite{Panfilo_2010}
\begin{equation}
    u = \frac{\sqrt{2}u_{\rm A}}{\tau_0}\left(\frac{\tau}{\tau_0} \right)^{-0.9},
\end{equation}
where $u_A = 0.2\,\mathrm{ns}$ is the satellite link phase instability for UTC(USNO) reported in the Circular T, $\tau_0=5\,$d, and here the measurement window $\tau$ is also 5 d. 

The resulting fractional deviations $y[f_\mathrm{Lu}/f_\mathrm{TAI}]$ for each 5 day window are shown in \fref{fig:measurements}(b) with all statistical uncertainty contributions summarized in \tref{ufor5days}.

\begin{table*}[h]
\centering
\footnotesize

\caption{\label{ufor5days} Statistical uncertainty contributions to $f_\mathrm{Lu}/f_\mathrm{TAI}$ for each 5 day window. }

\lineup
\begin{tabular}{llllllllll}
\toprule
MJD start                      & 59269 & 59274 & 59284 & 59289 & 59294 & 59314    & 59324   & 59329   & 59334                               \\
MJD stop                       & 59274 & 59279 & 59289 & 59294 & 59299 & 59319    & 59329   & 59334   & 59339                                \\ \midrule
Lu uptime                        & 11\%  & 2\%  & 7\%  & 37\%    & 14\%     & 9\%    & 3\%    & 30\% & 20\%                                \\ \midrule
Ratio&\multicolumn{9}{c}{Uncertainty $(\times 10^{-16}\,\mathrm{Hz/Hz})$}\\\midrule
Lu clock                             &\00.4 & \00.8 &\00.5 & \00.2 &\00.4    &\00.4 &\00.7&\00.2   &\00.3\\
HM interpolation                 & 15 & 29 & 16 &  \06.8  & 14 & 18    & 24   &  \07.6   & \08.8                                \\
HM/UTC(USNO)                   & 15 & 15 & 15& 15 & 15 & 15    & 15   & 15   & 15                                \\
UTC(USNO)/TAI                  & \06.6  & \06.6  & \06.6  & \06.6  & \06.6  & \06.6     & \06.6    & \06.6    & \06.6                             \\ \midrule
Lu/TAI Total                   & 22 & 33 & 23 & 18 & 21 & 24    & 29   & 18   & 19                               \\ 
\bottomrule
\end{tabular}
\end{table*}

\subsection{Link to SI } 
The measurement intervals span three Circular T reports; 399, 400, and 401. We evaluate the weighted averages of the 5-day results falling within each approximately monthly report interval. The uncertainty from interpolation of TAI over the 5-day dead time intervals is estimated by simulation, similar to the HM interpolation. The stability of TAI is derived from the free atomic timescale (Echelle Atomique Libre, EAL), and an instability model is reported for each Circular T~\cite{EAL}. For all three reports the model is given to be the sum of $1.4\times10^{-15}(\tau/\mathrm{s})^{-1/2}$ white frequency noise, $2\times10^{-16}$ flicker frequency noise, and $2\times10^{-17}(\tau/\mathrm{s})^{1/2}$ random walk frequency noise.

The Circular T reports the fractional deviation of the TAI scale interval relative the SI second as $d$ and its total uncertainty $u$. The estimation of $d$ is based on all primary and secondary frequency standard (PSFS) measurements contributing to TAI steering. Following the procedure reported by others~\cite{Pizzocaro_2020,hachisu2017absolute}, the total uncertainty $u$ is separated into a statistical component and systematic component using the systematic uncertainties of the individual PSFS reported as $\mu_B$ in the Circular T. For the three Circular T reports spanned, most of the clocks reporting are the same for all three months and so the total systematic component is assumed to be correlated. 
The uncertainty budget for each month and the final result are summarized in \tref{tab:si}.

\begin{table}
    \footnotesize
    \caption{\label{tab:si} Uncertainty budget for the absolute frequency measurement of the $^{176}\mathrm{Lu}^{+}$  $^{1}\rm S_0 - {^{3}}\rm D_1$ frequency standard for each month, and the total campaign.}
    \lineup
    \centering
    \begin{tabular}{lllll}
    
    \toprule
    &Cir.T&Cir.T& Cir.T& \\ 
    &399&400&401&Total\\ \midrule
     &\multicolumn{3}{c}{Statistical uncertainty $\mu_\mathrm{A}$ }\\
     &\multicolumn{3}{c}{$(\times 10^{-16}\,\mathrm{Hz/Hz})$ }\\
     Lu/TAI& \09.9&13&19&{\bf\07.3} \\ 
     TAI interpolation& \01.7&\02.9 &\06.3 &{\bf\01.5} \\  
     TAI/SI&\00.7 &\00.7 &\00.9 &{\bf\00.4}\\ 
     Total statistical& {\bf10}&{\bf13} &{\bf20}&{\bf\07.4} \\
     &\multicolumn{3}{c}{Systematic uncertainty $\mu_\mathrm{B}$}\\
     &\multicolumn{3}{c}{$(\times 10^{-16}\,\mathrm{Hz/Hz})$ }\\
     HM drift&\02.0&\02.0&\02.0&{\bf\02.0} \\
     HM diurnal&\05.0&\05.0&\05.0& {\bf\05.0}\\ 
     Lu sys. &\00.1&\00.1&\00.1&{\bf \00.1}\\
     Gravitational shift&\00.4 &\00.4 &\00.4 &{\bf\00.4}\\
     PSFS & \00.9&\00.8 &\00.9 &{\bf\00.9}\\
     Total systematic&{\bf\05.5}&{\bf\05.5}&{\bf\05.5}&{\bf\05.5} \\
     \\
     Lu/SI Total Unc.&{\bf11}&{\bf14}&{\bf20}&{\bf\09.2}\\ \bottomrule
    \end{tabular}
\end{table}

\begin{figure}[t]
    \centering
    \includegraphics[width=\columnwidth]{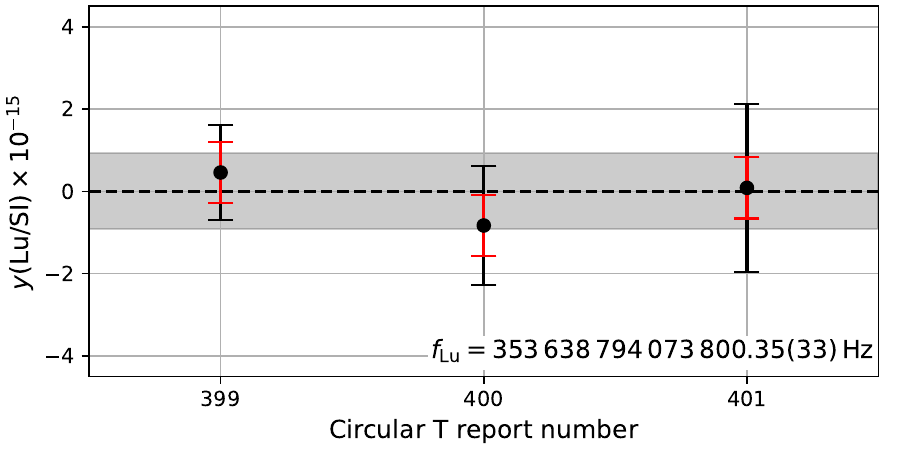}
    \caption{The measured frequency offset for each month (black points) relative to the weighted mean value. Black error bars are the total uncertainties, and red error bars the statistical component only.  The weighted mean of all three months is $353\,638\,794\,073\,800.35(33)\,$Hz, with the fractional uncertainty of $9.2\times10^{-16}$ indicated by the gray shaded region. }
    \label{fig:result}
 \end{figure}

\section{Discussion \label{sec:conclusion}}
The absolute frequencies obtained for each circular T report are shown in \fref{fig:result}. The result of $353\,638\,794\,073\,800.35(33)\,$Hz is obtained from the weighted mean of values from the three months.  The reduced chi squared is $\chi^2_\nu = 0.56$ with $\nu=2$ degrees of freedom.

The fractional uncertainty of $9.2\times10^{-16}$  is limited by the satellite link stability and HM interpolation uncertainty. For future campaigns, improved satellite link stability might be achieved by PPP with carrier-phase integer ambiguity resolution (PPP-AR)~\cite{Petit_2022,Jian_2023}. Processing of more recent GNSS data with the PPP-AR algorithm implemented by the Natural Resources Canada (NRCan) online service shows an instability of $\sim1\times10^{-15}$ at 1 d averaging time, compared to the 5 days observed in this work. To realize this improvement it was also necessary to relocate the rooftop GNSS aerial away from a nearby building partially obstructing the sky.  

The HM interpolation uncertainty may be reduced by either higher uptime of the optical clock or improved stability of the local frequency reference.  During preparation of this report, 95\% clock uptime was achieved over ten consecutive days. The primary source of clock interruptions remaining is mode hopping of a 701 nm laser which is frequency doubled to generate the 350 nm repump light.    An alternative solution for generating 350 nm light by third harmonic generation is in progress. If this proves a robust solution, uptime approaching 100\% appears to be imminently feasible, limited only by interruption due to ion loss. Our trapping lifetime for a single ion is typically several days in both experiment chambers.

A stabilized optical fiber link to the Singapore National Metrology Centre (NMC) has been recently established and will enable linking of our hydrogen masers. One of the hydrogen masers of NMC is specified to have  better stability than the one used here, but this is yet to be verified against our optical clock. An ensemble of hydrogen masers could provide an improved stability flywheel and better characterisation of the noise of the individual masers by continuous observation relative to the ensemble~\cite{nemitz2021absolute}. 

With these improvements an absolute frequency measurement at low $10^{-16}$ uncertainty is feasible. This would be comparable to state-of-the-art measurements based on satellite links to TAI~\cite{nemitz2021absolute,Pizzocaro_2020,kim2021absolute} and the accuracy of the caesium primary standards themselves.

\ack
%
The authors acknowledge Zhiqiang Zhang for contributions to initial data collection and curation. The authors thank Nils Nemitz for critical reading of the manuscript and Alex Voss for orthometric levelling measurements to determine the ellipsoidal height.

This research is supported by the National Research Foundation (NRF), Singapore, under its Quantum Engineering Programme (QEP-P5); the National Research Foundation, Singapore and A*STAR under its Quantum Engineering Programme (NRF2021-QEP2-01-P03) and through the National Quantum Office, hosted in A*STAR, under its Centre for Quantum Technologies Funding Initiative (S24Q2d0009); and the Ministry of Education, Singapore under its Academic Research Fund Tier 2 (MOE-T2EP50120-0014).

\section*{Data availability}

The RINEX files of USN7 were obtained through the online archives of the Crustal Dynamics Data Information System (CDDIS), NASA Goddard Space Flight Center, Greenbelt, MD, USA. \url{https://cddis.nasa.gov/archive/gnss/data/daily/}.

Data sets that support the findings of this study are available upon reasonable request from the corresponding author.

\section*{References}
\bibliographystyle{iopart-num}

\bibliography{biblio}

\end{document}